\documentclass{article}

\usepackage[final]{neurips_2023}

\usepackage[utf8]{inputenc}
\usepackage[T1]{fontenc}
\usepackage{hyperref}
\usepackage{url}
\usepackage{booktabs}
\usepackage{amsfonts}
\usepackage{nicefrac}
\usepackage{microtype}
\usepackage{xcolor}
\usepackage{bbding}
\usepackage{graphicx}
\usepackage{amsmath}
\usepackage{float}
\usepackage{multirow}
\usepackage{subfigure}
\usepackage{colortbl}
\usepackage{enumitem}

\newcommand{\Tref}[1]{Table~\ref{#1}}

\newcommand{\Fref}[1]{Figure~\ref{#1}}
\newcommand{\Sref}[1]{Section~\ref{#1}}

\newcommand{\ie}{\textit{i}.\textit{e}. }

\newcommand{\etal}{\textit{et al}. }

\title{Aparecium: Revealing Secrets from Physical Photographs}

\author{
  Zhe Lei \\
  Fudan University \\
  \texttt{zlei22@m.fudan.edu.cn} \\
  \And
  Jie Zhang \thanks{The corresponding author} \\
  Nanyang Technological University \\
  \texttt{jie\_zhang@ntu.edu.sg} \\
  \And
  Jingtao Li $^*$ \\
  Fudan University \\
  \texttt{lijt@fudan.edu.cn} \\
  \And
  Weiming Zhang \\
  \small{University of Science and Technology of China}  \\
  \texttt{zhangwm@ustc.edu.cn} 
  \And
  Nenghai Yu \\
  \small{University of Science and Technology of China} \\
  \texttt{ynh@ustc.edu.cn} \\
}

\begin{document}

\maketitle

\begin{figure}[h]
  \centering
  \includegraphics[width=1\linewidth]{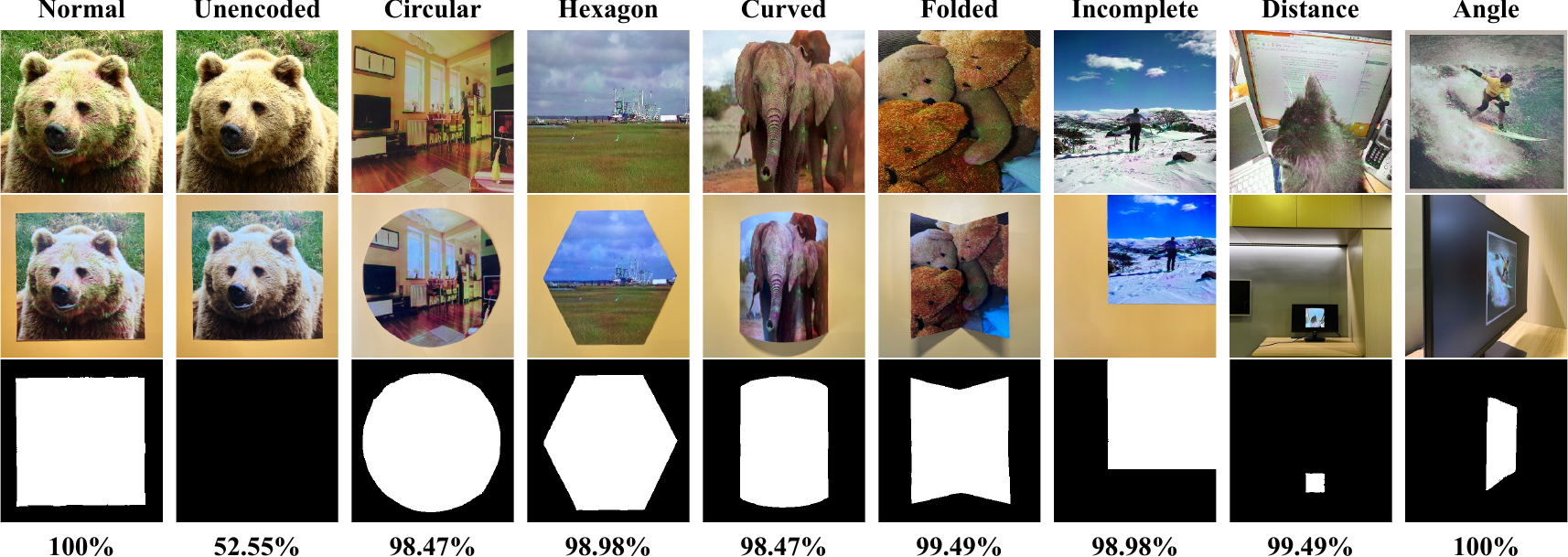}
  \caption{The performance of \textit{Aparecium} in the wild. We showcase different physical distortions in each column. The 2nd row is distorted images from printing-shooting or screen-shooting of the 1st row, respectively. The 3rd row displays the located watermarked areas, while the last row is the corresponding extraction accuracy.}
  \label{fig:in_the_wild_robustness}
\end{figure}

\begin{abstract}

Watermarking is a crucial tool for safeguarding copyrights and can serve as a more aesthetically pleasing alternative to QR codes. In recent years, watermarking methods based on deep learning have proved superior robustness against complex physical distortions than traditional watermarking methods. However, they have certain limitations that render them less effective in practice. For instance, current solutions necessitate physical photographs to be rectangular for accurate localization, cannot handle physical bending or folding, and require the hidden area to be completely captured at a close distance and small angle. To overcome these challenges, we propose a novel deep watermarking framework dubbed \textit{Aparecium}. Specifically, we preprocess secrets (\ie watermarks) into a pattern and then embed it into the cover image, which is symmetrical to the final decoding-then-extracting process. To capture the watermarked region from complex physical scenarios, a locator is also introduced. Besides, we adopt a three-stage training strategy for training convergence. Extensive experiments demonstrate that \textit{Aparecium} is not only robust against different digital distortions, but also can resist various physical distortions, such as screen-shooting and printing-shooting, even in severe cases including different shapes, curvature, folding, incompleteness, long distances, and big angles while maintaining high visual quality. Furthermore, some ablation studies are also conducted to verify our design.

\end{abstract}

\section{Introduction}

Watermarking is a widely-used technique for discreetly incorporating information into images without sacrificing the visual quality, while still allowing for information extraction despite various distortions. According to the type of embedded secrets, watermarking can be leveraged for different purposes. For example, embedding copyright information into images can be utilized for protection against copyright infringement, while embedding hyperlinks into images enables redirection to arbitrary information when the image is scanned using a mobile device, which can be regarded as an aesthetically pleasing alternative to unattractive QR codes. Robustness is the most significant objective for watermarking, making it effective in practical applications.

Traditional watermarking methods \cite{zhao1995embedding,voloshynovskiy2001multibit,hernandez2000dct,barni2001improved,ruanaidh1996phase} mainly focus on robustness against digital distortions, such as affine transformation, Gaussian noise, JPEG, etc. Commonly, each traditional method aims to resist a specific distortion. In recent years, deep learning-based watermarking methods \cite{zhu2018hidden, wengrowski2019light, tancik2020stegastamp, fang2020deep, fang2022pimog} have achieved tremendous success, which simulate the complex distortion in an end-to-end way that makes it easy to obtain general robustness. However, some of these methods, such as HiDDeN  \cite{zhu2018hidden}, are only resilient to digital distortions but fragile to physical distortions, where the image is physically captured. To address it, there are some following attempts. For example, Wengrowski and Dana \cite{wengrowski2019light} introduce a camera-display transfer function (CDTF) to obtain robustness against screen-shooting. Afterward, the popular StegaStamp \cite{tancik2020stegastamp} is proposed which can guarantee general robustness against both screen-shooting and print-shooting. Very recently, Jia \etal \cite{jia2022learning} propose Offline-to-online to embed watermarks into local regions, which can further improve the visual quality meanwhile preserve physical robustness. Nevertheless, there are still some limitations of StegaStamp \cite{tancik2020stegastamp} or Offline-to-online \cite{jia2022learning}: 1) it requires physical photographs to be rectangular for accurate localization; 2) it is unable to handle bending or folding, and 3) it necessitates the hidden area to be fully captured at a short distance and small angle. The above constraints degrade its effectiveness in the wild. 

To effectively reveal secrets from physical photographs, this paper presents a novel robust deep watermarking named \textit{Aparecium}, whose framework is displayed in \Fref{fig:framework}. The overall framework contains five modules, namely, a message processor, an encoder, a locator, a decoder, and a message extractor. After training, the former two modules and the last three modules are utilized for watermark embedding and extraction, respectively. Different from current methods \cite{wengrowski2019light,tancik2020stegastamp,jia2022learning} that map bit-string messages into a noise pattern, we propose to preprocess the bit-string message to a semantic pattern by a series of transposed convolutions, and then encode it into the target cover image, which is symmetrical to the subsequent decoding-then-extracting process. We point out that such incremental processing-then-encoding and  symmetrical decoding-then-extracting are easier for training. Besides, we also leverage a locator to help locate the watermarked area where the pattern potentially exists. More importantly, we introduce a three-stage training strategy to incrementally train the above-mentioned five modules. Specifically, at Stage $\textup{\uppercase\expandafter{\romannumeral1}}$, we jointly train the message processor and the message extractor, where some distortions are also introduced to enhance their robustness, such as random erasing, perspective transformation, and affine transformation. Finally, we can obtain the pattern to be embedded into the cover image.  At Stage $\textup{\uppercase\expandafter{\romannumeral2}}$, we fix the pre-trained message processor and the message extractor, and then jointly train the encoder and extractor to embed and extract the target pattern, respectively. In addition, the locator is also trained to locate the watermarked image. To guarantee both digital and physical robustness, spatial distortion, composing operation, and pixel-wise distortion are utilized. At Stage $\textup{\uppercase\expandafter{\romannumeral3}}$, we fine-tune all five modules for better performance.

Extensive experiments demonstrate that the proposed \textit{Aparecium} can achieve robustness against different digital and physical distortions meanwhile preserving satisfied visual quality. Importantly, \textit{Aparecium} is able to successfully reveal secrets from the target photograph suffering various physical distortions, including different shapes, curvature, folding, incompleteness, different distances, and different angles. We also conduct many ablation studies to verify our design. Finally, we point out some limitations that we will address in the future. 

Our contributions can be summarized as follows:

\begin{itemize}[leftmargin=*]

\item We propose a novel robust deep watermarking framework, \textit{Aparecium}, which can effectively reveal secrets from physical photographs.

\item For effective watermark extraction, we propose an incremental decoding-then-extracting manner rather than the widely-used direct extracting manner. Symmetrically, a processing-then-encoding manner is also leveraged for watermark embedding, where we adopt a series of transposed convolutions to effectively diffuse messages to the whole pattern. Besides, we introduce an extra locator for localizing the watermarked region of physical photographs.

\item To achieve superior robustness and training convergence, we adopt a three-stage training strategy.
we leverage spatial distortions, composing operation, and pixel-wise distortions to enhance both digital robustness and physical robustness.

\item Extensive experiments demonstrate the proposed \textit{Aparecium} can achieve both digital and physical robustness, including some severe scenarios such as long shooting distances and big shooting angles. Besides, we also execute many ablation studies to prove the effectiveness of our design, which we want to shed some light on the field of deep watermarking.

\end{itemize}

\section{Related Work}

\subsection{Digital Robust Deep Watermarking}

Digital robust deep watermarks are particularly suitable for scenarios where images are transmitted in the digital channel. Baluja \cite{baluja2017hiding} proposed the first end-to-end training framework for information hiding and extraction. Based on it, HiDDeN \cite{zhu2018hidden} appended a noise layer between watermark embedding and extraction, which included a series of differentiable distortions, thereby obtaining the desired robustness to such distortions. Importantly, the non-differentiable like JPEG can be approximated by differentiable operations. For enhancing robustness to real JPEG, MBRS \cite{jia2021mbrs} proposed randomly switching between real JPEG, simulated JPEG, and noise-free layers in each mini-batch. Then, Luo \etal \cite{luo2020distortion} leveraged adversarial training to improve robustness against unknown distortions. 
To reduce the time of watermark embedding, Zhang \etal \cite{zhang2020udh} proposed a novel framework named UDH. Moreover, some recent methods \cite{lu2021large, jing2021hinet, guan2022deepmih, xu2022robust} also explore the effectiveness of invertible neural networks for watermark embedding and extracting. 
However, all these methods are fragile to physical distortions such as screen-shooting and print-shooting.

\subsection{Physical Robust Deep Watermarking}

Our approach is closely related to physical robust deep watermarking, which not only ensures robustness for digital distortions but also for physical distortions. Previous studies, such as LFM \cite{wengrowski2019light} and PIMoG \cite{fang2022pimog}, have focused on enhancing the robustness against screen shooting. LFM \cite{wengrowski2019light} achieved robustness by training a CDTF network, while PIMoG \cite{fang2022pimog} designed the most influenced distortion in the noise layer. Both methods simulated the distortion present in screen-shooting and achieved excellent robustness. 
However, their robustness against print-shooting is limited, namely, they don't optimize for print-shooting and lack a localization network to locate the watermark area in the photograph.

In contrast, StegaStamp \cite{tancik2020stegastamp} and Offline-to-online \cite{jia2022learning} have achieved robustness against both screen-shooting and print-shooting. Similar to HiDDeN \cite{zhu2018hidden}, StegaStamp \cite{tancik2020stegastamp} simulates these physical distortions into the training process, which sacrifices visual quality for certain robustness. To improve visual quality, Offline-to-online \cite{jia2022learning} hides information in sub-images, namely, a local region of original images, and requires a localization network to locate the watermarked region. However, during the watermark extracting, both methods require physical photographs to be presented in a complete and flat manner during shooting, which is not applicable in some practical scenarios. In this paper, we adopt StegaStamp \cite{tancik2020stegastamp} and Offline-to-online \cite{jia2022learning} as the baseline methods.

\begin{figure}[t]
  \centering
  \includegraphics[width=1\linewidth]{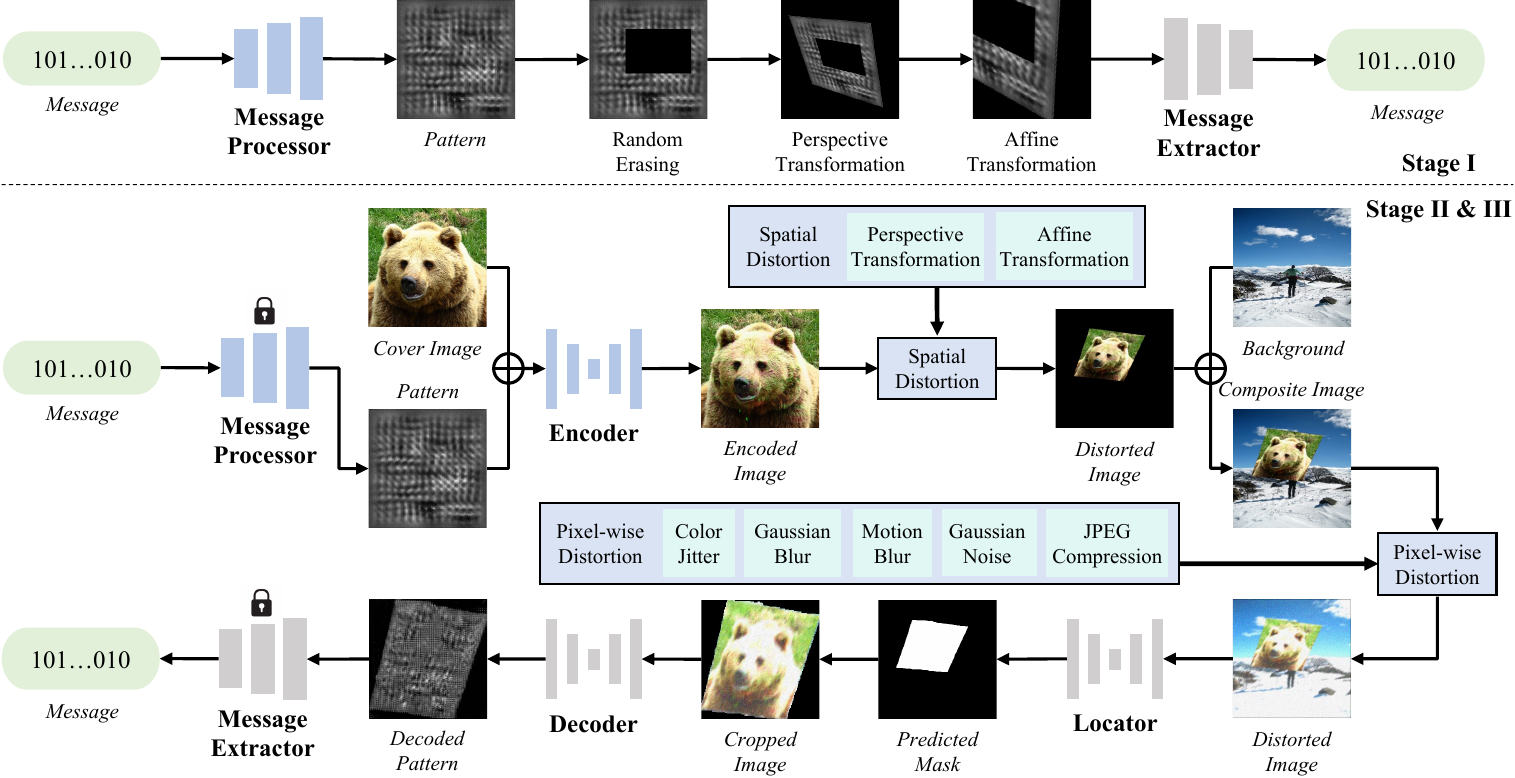}
  \caption{The overall framework of the proposed method, where we adopt a three-stage training strategy. At Stage $\textup{\uppercase\expandafter{\romannumeral1}}$, we jointly train the message processor and the message extractor, which are fixed at Stage $\textup{\uppercase\expandafter{\romannumeral2}}$ while we optimize all models at Stage $\textup{\uppercase\expandafter{\romannumeral3}}$. After training, the blue-colored models are used for watermark embedding and the gray-colored models are leveraged for watermark extraction.}
  \label{fig:framework}
\end{figure}

\section{Methodology}

\subsection{The Framework of \textit{Aparecium}}

\paragraph{Overview.}
As shown in \Fref{fig:framework}, \textit{Aparecium} consists of several components, including a message processor, an encoder, a locator, a decoder, and a message extractor. Given a message, the message processor transforms it into a pattern, which is then encoded into a cover image by the encoder while maintaining visual similarity. The resulting encoded image can be printed or displayed on a screen and subsequently captured. The locator component then identifies the position of the encoded image and generates a mask that is used to automatically crop out the encoded image. The pattern decoder then decodes the pattern, and the hidden message is ultimately extracted by the message extractor based on the recovered pattern. In the following sections, 
we will describe each component in detail.

\paragraph{Message Processor.}
The message processor component is responsible for converting the message into a single-channel pattern, which enables the encoder to conceal the message within the cover image more effectively. To achieve better robustness, the message processor aims to distribute the message as evenly as possible throughout the pattern. The message is represented as a 196-bit binary string, which is concatenated along the channel dimension to create a 196$\times$1$\times$1 tensor. This tensor is then upsampled using a series of transposed convolutions, resulting in a single-channel pattern with a size or dimension of 256$\times$256.

\paragraph{Encoder.}
The encoder component aims to encode the single-channel pattern into a three-channel RGB image while minimizing the visual discrepancies between the encoded image and the cover image. Specifically, we employ the widely-used U-Net \cite{ronneberger2015u} as its architecture, which accepts a 256$\times$256 image with four channels (three-channel RGB cover image plus a single-channel pattern) and outputs a three-channel RGB encoded image.

\paragraph{Locator.}
The locator is introduced to identify the location of the encoded image within a captured physical photograph. To address situations such as incomplete image capture, we have selected the salient object detection network U$^2$-Net \cite{qin2020u2} as our locator. For any resolution photo, we resize it to 320$\times$320. The locator then identifies the position of the encoded image and generates a 320$\times$320 mask to depict the region of the encoded image.

\paragraph{Decoder.}
After the location of the encoded image has been identified by the locator, the image is automatically cropped based on the generated mask and resized to 256$\times$256. The main function of the decoder is to decode the single-channel pattern from the cropped image. During the pattern decoding process, pixel-wise distortions such as color distortion, noise, and blur are randomly introduced to enhance the robustness against such distortions. We employ a U-Net \cite{ronneberger2015u} as its architecture. In a nutshell, the decoder takes a 256$\times$256 cropped image as input and generates a single-channel 256$\times$256 decoded pattern as output.

\paragraph{Message Extractor.}
The message extractor is a critical component in retrieving hidden messages from a given pattern. However, even after the message has been decoded, the pattern may still exhibit spatial distortions, such as perspective and affine transformations. To address this issue, we utilized the ConvNeXt \cite{liu2022convnet} classification network as our message extractor. This network accepts a 256$\times$256 pattern as input and generates a tensor of length 196, namely the final extracted bit string.

\subsection{Three-stage Training Strategy}
To achieve the desired functionality of the above five modules, we adopt a three-stage training strategy, which will be introduced in the following part.

\paragraph{Training Stage $\textup{\uppercase\expandafter{\romannumeral1}}$.}
In the first stage, our focus is on training the message processor and message extractor. Specifically, the message processor is responsible for converting randomly generated binary strings into patterns. To ensure that the message is diffused into the pattern and that the message extracting process can withstand spatial distortions, we introduce various distortions to the pattern prior to extracting. These distortions include (a) random erasing, which involves randomly erasing a rectangular portion of the image; (b) perspective transformation, which simulate a scenario where the camera and the image are not aligned; and (c) affine transformation, which encompass rotation, translation, and scaling. Ultimately, the message extractor is able to extract the hidden message from the distorted image. We adopt $L_{\textup{\uppercase\expandafter{\romannumeral1}}}$ to constrain training stage $\textup{\uppercase\expandafter{\romannumeral1}}$, \ie,

\begin{equation}
\begin{aligned}
  L_{\textup{\uppercase\expandafter{\romannumeral1}}} = L_{msg} = BCE(m, m'),
\end{aligned}
\end{equation}
where $m$ is the ground truth message and $m'$ is the extracted message.

\paragraph{Training Stage $\textup{\uppercase\expandafter{\romannumeral2}}$.}
In this stage, we fixed the parameters of the message processor and message extractor and only train the encoder, locator, and decoder. Once the message processor generates a pattern based on the random message, the encoder encodes the pattern and localization information into the cover image. To simulate spatial distortion that may occur during the photo-taking process, we apply spatial distortions to the encoded image, including perspective transformation and affine transformation. Additionally, we also apply spatial distortions to the pattern and generate the ground truth mask for the subsequent process. After composing the spatially distorted image with the background image, we added pixel-wise distortions to the composite images, including brightness, contrast, saturation, hue, Gaussian blur, motion blur, Gaussian noise, and JPEG compression. Next, the locator locates the position of the watermarked region of the distorted composite image. Then, we will obtain the input of the following decoder and ground truth pattern after cropping. Finally, the decoder decodes the pattern from the cropped image. The loss function for stage $\textup{\uppercase\expandafter{\romannumeral2}}$ consists of visual loss $L_{vis}$, localization loss $L_{loc}$, and pattern loss $L_{pattern}$, \ie,

\begin{equation}
\begin{aligned}
  L_{\textup{\uppercase\expandafter{\romannumeral2}}} & = \lambda_1 L_{vis} + \lambda_2 L_{loc} + \lambda_3 L_{pattern} \\
  &= \lambda_1 (MSE(I_{co}, I_{en}) +  SSIM(I_{co}, I_{en})) + \lambda_2 BCE(M_{gt}, M_{pr}) \\
  & \;\;\;\; + \lambda_3 (MSE(P_{gt}, P_{de}) + SSIM(P_{gt}, P_{de})),
\end{aligned}
\end{equation}

where $I_{co}$ refers to the cover image, $I_{en}$ represents the encoded image, $M_{gt}$ denotes the ground truth mask, $M_{pr}$ indicates the predicted mask, $P_{gt}$ stands for the ground truth pattern, and $P_{de}$ represents the decoded pattern.

\paragraph{Training Stage $\textup{\uppercase\expandafter{\romannumeral3}}$.}
During the third stage, we fine-tune all five modules in an end-to-end way. The loss function $L_{\textup{\uppercase\expandafter{\romannumeral3}}}$ can be written as follows:

\begin{equation}
\begin{aligned}
  L_{\textup{\uppercase\expandafter{\romannumeral3}}} & = \lambda_1 L_{vis} + \lambda_2 L_{loc} + \lambda_3 L_{pattern} + \lambda_4 L_{msg} \\
  & = \lambda_1 (MSE(I_{co}, I_{en}) +  SSIM(I_{co}, I_{en})) + \lambda_2 BCE(M_{gt}, M_{pr}) \\
  & \;\;\;\; + \lambda_3 (MSE(P_{gt}, P_{de}) + SSIM(P_{gt}, P_{de})) + \lambda_4 BCE(m, m').
\end{aligned}
\end{equation}

\section{Experiment}

\subsection{Experiment Settings}

\paragraph{Dataset.}
During the training phase, we utilize the COCO 2017 training dataset and compose a message consisting of 196 randomly generated binary bits. For the testing phase, we select 100 images from the COCO 2017 validation dataset.

\paragraph{Metrics.}
The evaluation of image visual quality is conducted using two metrics: Peak Signal to Noise Ratio (PSNR) and Structural Similarity Index (SSIM). The decoding performance is evaluated using Bit Error Ratio (BER).

\begin{table}[t]
  \caption{Compare with the state-of-the-arts methods.}
  \centering
  \setlength{\tabcolsep}{23.9pt}
  \begin{tabular}{ccccc}
    \toprule
    Method & PSNR & SSIM & Manual Cropped \\
    \midrule
    StegaStamp \cite{tancik2020stegastamp} & 26.46 & 0.8802 & \XSolidBrush \\
    Offline-to-online \cite{jia2022learning}  & 31.01 & 0.9648 & \Checkmark \\
    Ours & 33.48 & 0.9883 & \XSolidBrush \\
    \bottomrule
  \end{tabular}
  \label{tab:visual_quality}
\end{table}

\begin{figure}[t]
  \centering
  \includegraphics[width=1\linewidth]{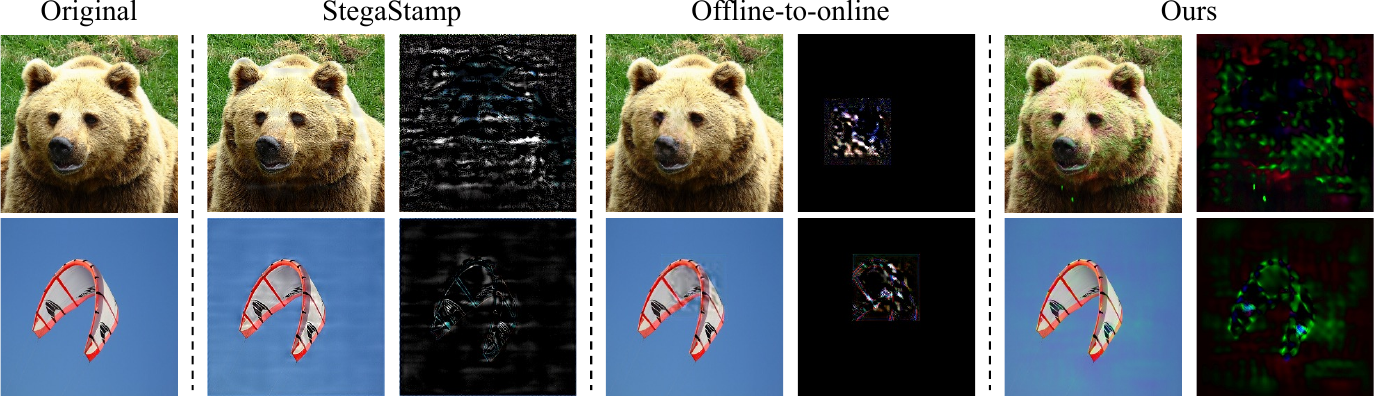}
  \caption{Visual examples of different watermarking methods. The 1st column is original watermark-free images. For each method, the 1st column and 2nd column are watermarked images, and the corresponding 5$\times$ residual between watermark-free and watermarked images, respectively. }
  \label{fig:visual_quality}
\end{figure}

\paragraph{Implementation Details.}
We compare \textit{Aparecium} with the state-of-the-art methods, \ie, StegaStamp \cite{tancik2020stegastamp} and Offline-to-online \cite{jia2022learning}. To ensure a fair comparison, we retrain StegaStamp to have the same image size (256$\times$256) and message length (196 bits). During the physical world experiment, we utilize a DELL S2421NX monitor for display and use iPhone 13 Pro for shooting by default. The images are displayed at their default size. Other devices such as a MacBook Pro14 laptop and a Redmi 12C are also leveraged for justifying the general robustness. The default shooting angle is 0 $^\circ$, while the default distance for screen-shooting and print-shooting are 30cm and 90cm, respectively. More implementation details can be found in supplementary materials.

\subsection{Visual Quality}
As shown in \Tref{tab:visual_quality}, \textit{Aparecium} outperforms other methods in terms of visual quality. The reason why the visual quality of Offline-to-online \cite{jia2022learning}  is lower than what they report (PSNR: 31.01 v.s. 32.95; SSIM: 0.9648 v.s. 0.9677) is that we randomly select sub-image positions for a fair comparison, instead of manually selecting high-frequency areas. Some visual examples are showcased in \Fref{fig:visual_quality}. Moreover, both StegaStamp and \textit{Aparecium} can automatically locate and crop the physical photographs, whereas Offline-to-online \cite{jia2022learning} requires manual cropping before locating the sub-image, making it inefficient in practice.

\begin{figure}[t]
  \centering
  \includegraphics[width=1\linewidth]{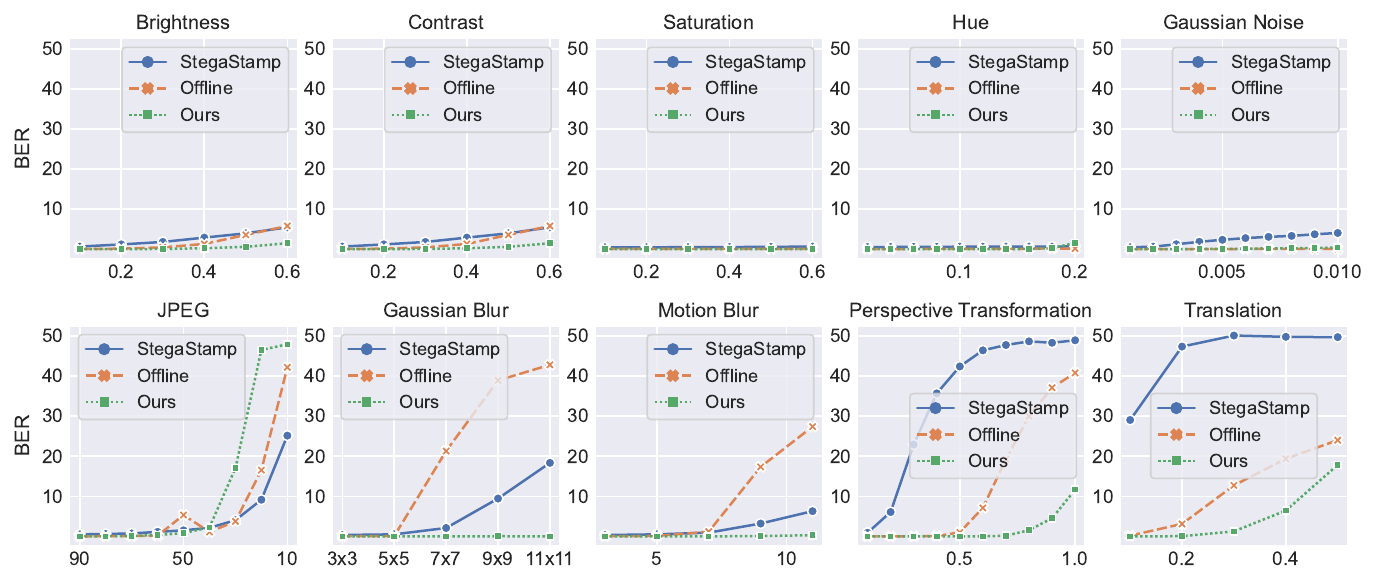}
  \caption{Comparison of robustness against different digital distortions.}
  \label{fig:digital_robustness}
\end{figure}

\begin{table}[t]
  \caption{Comparison of robustness against screen-shooting under different shooting distances.}
  \centering
  \setlength{\tabcolsep}{15.1pt}
  \begin{tabular}{cccccc}
    \toprule
    Distance & 10cm & 20cm & 30cm & 40cm & 50cm \\
    \midrule
        Offline-to-online \cite{jia2022learning} & 0.28$\%$ & 0.34$\%$ & 0.92$\%$ & 4.22$\%$ & 3.21$\%$ \\
    Ours & 0.72$\%$ & 0.33$\%$ & 0.76$\%$ & 2.23$\%$ & 1.73$\%$ \\
    \bottomrule
  \end{tabular}
  \label{tab:screen_shooting_distances}
\end{table}

\begin{table}[!ht]
  \caption{Comparison of robustness against screen-shooting under different shooting angles.}
  \centering
  \setlength{\tabcolsep}{7.3pt}
  \begin{tabular}{ccc|ccc}
    \toprule
    Horizontal & Offline-to-online \cite{jia2022learning} & Ours & Vertical & Offline-to-online \cite{jia2022learning} & Ours \\
    \midrule
    Left 60$^\circ$ & 6.16$\%$ & 1.75$\%$ & Up 60$^\circ$ & 5.00$\%$ & 1.99$\%$ \\
    Left 45$^\circ$ & 2.95$\%$ & 1.55$\%$ & Up 45$^\circ$ & 1.22$\%$ & 0.65$\%$ \\
    Left 30$^\circ$ & 0.91$\%$ & 0.52$\%$ & Up 30$^\circ$ & 0.94$\%$ & 0.46$\%$ \\
    0$^\circ$ & 1.16$\%$ & 0.41$\%$ & 0$^\circ$ & 1.16$\%$ & 0.41$\%$ \\
    Right 30$^\circ$ & 0.81$\%$ & 0.19$\%$ & Down 30$^\circ$ & 1.94$\%$ & 0.49$\%$ \\
    Right 45$^\circ$ & 2.40$\%$ & 0.55$\%$ & Down 45$^\circ$ & 3.04$\%$ & 1.09$\%$ \\
    Right 60$^\circ$ & 1.48$\%$ & 0.53$\%$ & Down 60$^\circ$ & 3.93$\%$ & 1.83$\%$ \\
    \bottomrule
  \end{tabular}
  \label{tab:screen_shooting_angles}
\end{table}

\begin{table}[t]
  \caption{Comparison of robustness against print-shooting under different shooting distances.}
  \centering
  \setlength{\tabcolsep}{14.9pt}
  \begin{tabular}{cccccc}
    \toprule
    Distance & 30cm & 60cm & 90cm & 120cm & 150cm \\
    \midrule
    Offline-to-online \cite{jia2022learning} & 0.33$\%$ & 0.46$\%$ & 1.06$\%$ & 3.57$\%$ & 4.23$\%$ \\
    Ours & 0.20$\%$ & 0.32$\%$ & 1.08$\%$ & 1.48$\%$ & 3.96$\%$ \\
    \bottomrule
  \end{tabular}
  \label{tab:print_shooting_distances}
\end{table}

\begin{table}[t]
  \caption{Comparison of robustness against print-shooting under different shooting angles.}
  \centering
  \setlength{\tabcolsep}{7.3pt}
  \begin{tabular}{ccc|ccc}
    \toprule
    Horizontal & Offline-to-online \cite{jia2022learning} & Ours & Vertical & Offline-to-online \cite{jia2022learning} & Ours \\
    \midrule
    Left 60$^\circ$ & 6.75$\%$ & 4.28$\%$ & Up 60$^\circ$ & 9.89$\%$ & 3.85$\%$ \\
    Left 45$^\circ$ & 6.18$\%$ & 2.97$\%$ & Up 45$^\circ$ & 4.20$\%$ & 1.77$\%$ \\
    Left 30$^\circ$ & 3.15$\%$ & 1.29$\%$ & Up 30$^\circ$ & 2.93$\%$ & 1.45$\%$ \\
    0$^\circ$ & 0.87$\%$ & 0.75$\%$ & 0$^\circ$ & 0.87$\%$ & 0.75$\%$ \\
    Right 30$^\circ$ & 2.35$\%$ & 0.82$\%$ & Down 30$^\circ$ & 1.02$\%$ & 1.13$\%$ \\
    Right 45$^\circ$ & 4.68$\%$ & 1.39$\%$ & Down 45$^\circ$ & 2.92$\%$ & 0.72$\%$ \\
    Right 60$^\circ$ & 9.83$\%$ & 2.54$\%$ & Down 60$^\circ$ & 6.40$\%$ & 4.07$\%$ \\
    \bottomrule
  \end{tabular}
  \label{tab:print_shooting_angles}
\end{table}

\begin{table}[!ht]
  \caption{The decoding results under different combinations of display and shooting devices.}
  \centering
  \setlength{\tabcolsep}{35.6pt}
  \begin{tabular}{ccc}
    \toprule
    Device & MacBook Pro 14 & Dell S2421NX \\
    \midrule
    iPhone 13 pro & 0.45$\%$ & 0.40$\%$ \\
    Redmi 12C & 1.43$\%$ & 1.29$\%$ \\
    \bottomrule
  \end{tabular}
  \label{tab:different_devices}
\end{table}

\begin{table}[!ht]
  \caption{Some robustness in the wild.}
  \centering
  \setlength{\tabcolsep}{20pt}{
  \begin{tabular}{ccccc}
    \toprule
    Distortion & Circular & Hexagon & Curved & Folded \\
    \midrule
    BER & 0.36$\%$ & 0.66$\%$ & 4.08$\%$ & 0.77$\%$ \\
    \bottomrule
  \end{tabular}}
  \label{tab:in_the_wild_robustness}
\end{table}

\subsection{Robustness Against Digital Distortions}

To compare the digital robustness with the baseline methods, we simulate various distortions and present the results in \Fref{fig:digital_robustness}. All methods achieve excellent results in terms of brightness, contrast, saturation, hue, and Gaussian noise. However, \textit{Aparecium} obtains lower robustness to JPEG compression, which can be improved by adjusting hyper-parameter $\lambda_{1}$ (shown \Sref{sec:hy}). For Gaussian blur and motion blur, \textit{Aparecium} demonstrates superior robustness compared to the other methods. More importantly, \textit{Aparecium} exhibits exceptional performance in spatial distortion, including perspective transformation and translation. The success of Offline-to-online \cite{jia2022learning} decoding under translation distortion depends on whether the sub-image is fully presented.

\begin{figure}[!ht]
  \centering
  \includegraphics[width=1\linewidth]{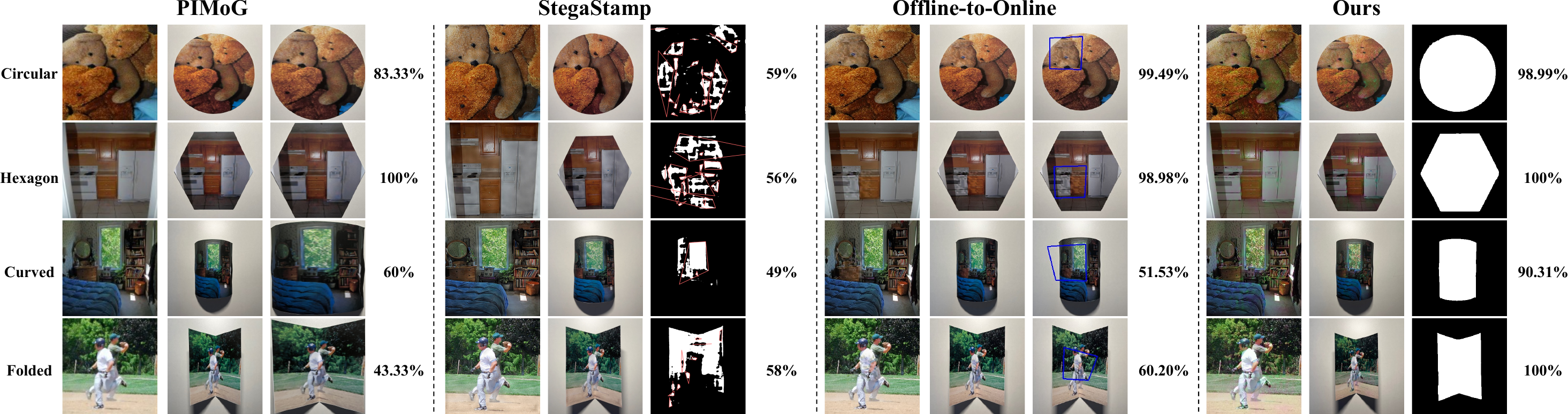}
  \caption{A visual example of comparison with the baselines about robustness in the wild. For each method, three columns represent watermarked images, processed watermarked images, and auxiliary strategy for extraction (\ie, calibration for PIMoG and localization for the other methods).}
  \label{fig:in_the_wild_robustness2}
\end{figure}

\subsection{Robustness Against Physical Distortions}

In this experiment, we evaluate the decoding performance of screen and print shooting at varying distances and angles. To ensure fairness, we compare our method with Offline-to-online \cite{jia2022learning} under fixed lighting conditions. For screen shooting, we display the image on a monitor with dimensions of 7cm $\times$ 7cm and capture photos by mobile phone. The results for different distances and angles are presented in \Tref{tab:screen_shooting_distances} and \Tref{tab:screen_shooting_angles}, and our method achieves a lower BER than Offline-to-online \cite{jia2022learning}. However, at a distance of 40cm, we encounter severe moire patterns that affect both our method and the baseline. Next, we test print shooting at different distances and angles and print them with dimensions of 20cm $\times$ 20cm. As shown in \Tref{tab:print_shooting_distances} and \Tref{tab:print_shooting_angles}, our method still achieves better performance.
Moreover, \Tref{tab:different_devices} shows the general robustness of our method on different devices.

\subsection{Robustness in the Wild}

To demonstrate the unexpected robustness of \textit{Aparecium} in real-world scenarios, we capture a series of photographs using a handheld mobile phone. The \Fref{fig:in_the_wild_robustness} showcases examples of the captured images with masks and decoding accuracy. In the 1st and 2nd columns, we capture both encoded and unencoded images, respectively. The results suggest that the locator works actually based on the embedded information rather than differences between the image and the background. According to the other results, \textit{Aparecium} performs well in the wild, even in some severe cases, such as bending, folding, incomplete capture, long shooting distances, big shooting angles, etc.

\begin{table}[t]
  \caption{Performance with different message processors.}
  \centering
  \setlength{\tabcolsep}{21.6pt}{
  \begin{tabular}{cc|ccc}
    \toprule
    Processor & Pattern BER & PSNR & SSIM &  BER \\
    \midrule
    Binary & 15.70$\%$ & 31.37 & 0.9801 & 6.68$\%$ \\
    LECA & 35.14$\%$ & 33.06 & 0.9938 & 43.37$\%$ \\
    StegaStamp & 2.71$\%$ & 32.06 & 0.9879 & 6.28$\%$ \\
    Ours & 2.01$\%$ & 33.48 & 0.9883 & 5.03$\%$ \\
    \bottomrule
  \end{tabular}}
  \label{tab:different_processors}
\end{table}

\begin{table}[t]
  \caption{Compare with different Message Extractors.}
  \centering
  \setlength{\tabcolsep}{33.8pt}
  \begin{tabular}{cccc}
    \toprule
    Extractor & PSNR & SSIM & BER \\
    \midrule
    ResNet \cite{he2016deep} & 32.01 & 0.9897 & 6.79$\%$ \\
    ConvNeXt & 33.48 & 0.9883 & 5.03$\%$ \\
    \bottomrule
  \end{tabular}
  \label{tab:different_extractors}
\end{table}

\begin{figure}[!ht]
  \centering
  \includegraphics[width=0.9\linewidth]{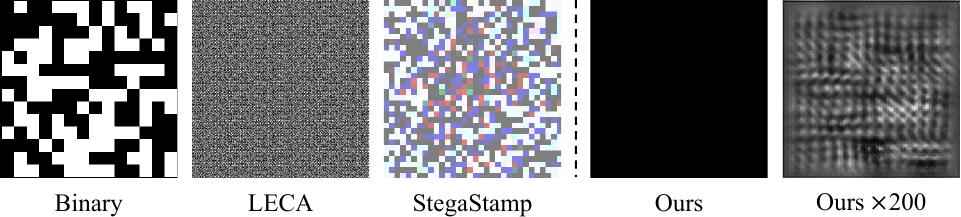}
  \caption{Output patterns from different Message Processors.}
  \label{fig:pattern}
\end{figure}

\begin{figure}[!ht]
  \centering
  \includegraphics[width=0.6\linewidth]{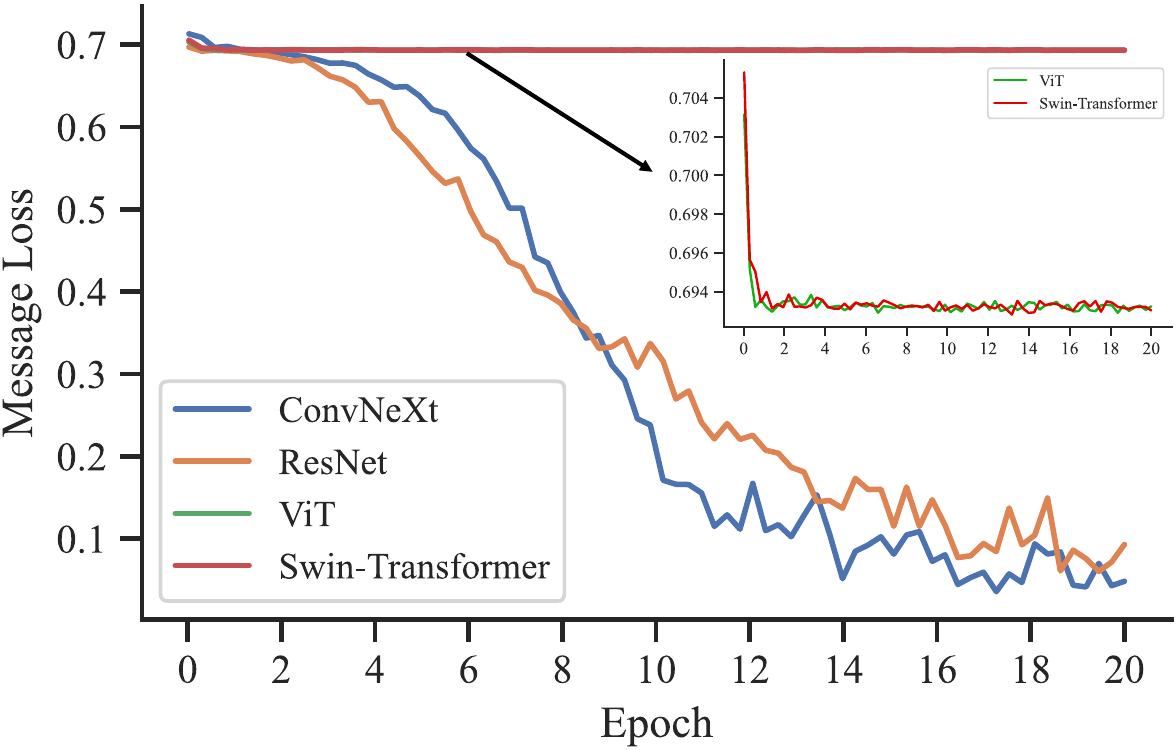}
  \caption{Training loss of different message extractors at Stage $\textup{\uppercase\expandafter{\romannumeral1}}$.}
  \label{fig:different_extractors}
\end{figure}

We include a comparison with other methods in the wild scenarios. As depicted in \Fref{fig:in_the_wild_robustness2}, StegaStamp fails to extract information accurately due to its inability to precisely locate the watermark region in the photograph. Offline-to-Online performs well in circular and hexagonal images as the watermark sub-images remain intact. However, it struggles to successfully extract information in cases involving bending and folding, primarily due to inaccurate positioning. In the case of PIMoG, manual perspective transformation is required to correct the images as this method lacks a locator. Once the correction is applied, information can be extracted from images in circular and hexagonal cases, but it is unable to handle curved and folded scenarios. Furthermore, we have supplemented \Tref{tab:in_the_wild_robustness} with quantitative data on the robustness in real-world scenarios.

\subsection{Ablation Study}

\paragraph{The Influence of Different Message Processors.}
To facilitate the embedding of bit information into images, many methods employ a processor to transform the information into patterns. LFM \cite{wengrowski2019light}, UDH \cite{zhang2020udh}, Offline-to-Online \cite{jia2022learning}, and other methods process the information into binary patterns, where white blocks represent 1 and black blocks represent 0. LECA \cite{luo2022leca} processes the information through a fully connected layer, then repeats spatially to match the dimensions of the cover image. StegaStamp, after passing through a fully connected layer, upsamples the pattern to match the dimensions of the cover image. In our method, we utilize a sequence of transposed convolutions as the message processor. As shown in \Fref{fig:pattern}, the visual message pattern of our method is more invisible. 

To evaluate the impact of this component, firstly, we train the above processors in an end-to-end way as the first stage, and we use "Pattern BER" to evaluate the performance in the first stage. Then, we train the subsequent stages of these processors as the proposed method and test the corresponding "PSNR", "SSIM", and "BER". As shown in \Tref{tab:different_processors}, our method outperforms the compared methods.

\paragraph{The Influence of Different Message Extractors.}
We employ ConvNeXt \cite{liu2022convnet} as the message extractor in our study. To examine the influence of different extractors, we further try classic ResNet \cite{he2016deep}, and transformer-based classifiers such as ViT \cite{dosovitskiy2020image} and Swin-Transformer \cite{liu2021swin}. As shown in the \Fref{fig:different_extractors} , ResNet extractor can still work while transformer-based classifiers fail to converge. We further test the impact of different extractors on visual quality and robustness, and the results are shown in \Tref{tab:different_extractors}.

\begin{figure}[t]
  \centering
  \includegraphics[width=1\linewidth]{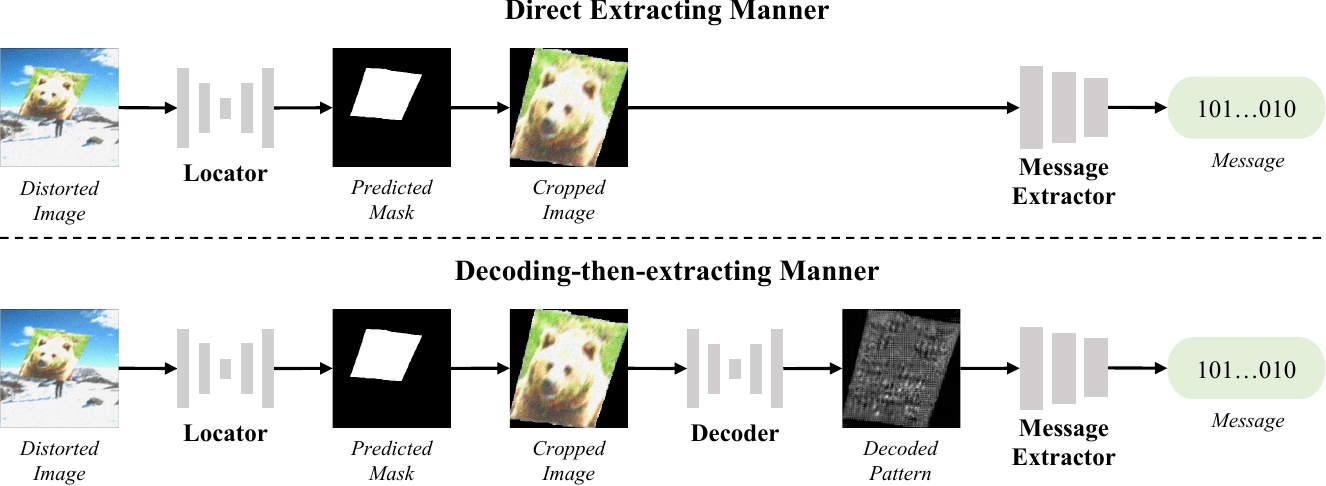}
  \caption{Comparison of training loss between the decoding-then-extracting manner and the direct extracting manner.}
  \label{fig:decoding_then_extracting}
\end{figure}

\begin{table}[t]
\caption{The influence of $\lambda_{1}$.}
  \centering
  \setlength{\tabcolsep}{10.4pt}
  \begin{tabular}{cccccccc}
    \toprule
    \multirow{2}{*}{Method} & \multirow{2}{*}{PSNR} & \multirow{2}{*}{SSIM} & \multirow{2}{*}{BER} & \multicolumn{4}{c}{JPEG} \\
    \cmidrule(lr){5-8}
     & & & & 40 & 30 & 20 & 10 \\
    \midrule
    $\lambda_{1}=1$ & 26.14 & 0.9619 & 1.36$\%$ & 0.00$\%$ & 0.00$\%$ & 1.57$\%$ & 44.62$\%$ \\
    $\lambda_{1}=5$ & 30.02 & 0.9800 & 3.76$\%$ & 0.00$\%$ & 2.30$\%$ & 10.02$\%$ & 47.28$\%$ \\
    $\lambda_{1}=10$ & 33.48 & 0.9883 & 5.03$\%$ & 2.26$\%$ & 17.05$\%$ & 46.54$\%$ & 47.96$\%$ \\
    \bottomrule
  \end{tabular}
  \label{tab:different_lambda1}
\end{table}

\paragraph{The Effectiveness of the Incremental Decoding-then-extracting.}

As show in \Fref{tab:different_lambda1}, current deep watermarking usually directly extracts the bit-string message from the watermarked image. In comparison, we adopt a decoding-then-extracting manner to extract messages, namely, using a decoder to reveal patterns from captured images and then a message extractor to extract messages from the decoded patterns. To demonstrate the superiority of our incremental manner, we also try the direct extracting manner but ultimately fail to achieve convergence due to its weaker extraction capability.

\paragraph{The Influence of Hyper-parameter $\lambda_{1}$.}
\label{sec:hy}
During the training process in the third stage, we set $\lambda_{1}=10$. We also train models with $\lambda_{1}=5$ and $\lambda_{1}=1$ to investigate the impact of $\lambda_{1}$. As shown in the \Tref{tab:different_lambda1}, lower $\lambda_{1}$ leads to poorer visual quality but better extraction and robustness. 

\paragraph{Replace the portion of StegaStamp with ours.}
We apply pixel-wise distortion, spatial distortion, and blending operation from our method to StegaStamp. However, after training, the modified StegaStamp model failed to converge. We suspect that the StegaStamp model is more vulnerable to spatial distortion compared to our method, which ultimately leads to convergence issues.

We replace the Message Processor of StegaStamp with ours, but the results still fail to converge. We suspect that this is due to different training strategies. The training strategy of StegaStamp may not sufficiently diffuse the message into the pattern. Additionally, our message processor is more complex than StegaStamp's, and its weaker decoding capability makes it difficult to achieve convergence.

\section{Conclusion}

This paper proposes a novel deep watermarking framework, \textit{Aparecium}, which is well-suited for real-world applications. Specifically, we investigate an efficient approach for processing watermark information and opt to use a series of transposed convolutions to diffuse the message into a pattern. In the decoding process, we distinguish pixel-wise distortion from spatial distortion, which significantly enhances the decoding performance. Finally, we adopt a three-stage training strategy to ensure training stability. Experiments demonstrate the proposed method achieves remarkable robustness against both digital distortions and physical distortions while preserving satisfied visual quality.

\section*{Limitations and Broader Impact}

We find that embedding messages into dark images may produce bright spots, because of their low redundancy for watermarking. Besides, when confronted with pronounced moire patterns, the performance will degrade. The related results can be found in supplementary materials. 

This work is of importance in preventing copyright infringement. Moreover, embedding hyperlinks into images can be used for information sharing by scanning photographs, which can be regarded as an aesthetically pleasing alternative to unattractive QR codes.

{
    \small
    \bibliographystyle{unsrt}
    \bibliography{ref}
}

\clearpage
\setcounter{section}{0}

\section{Implementation Details}
In this section, we describe the implementation of the proposed method in detail. Specifically, we showcase training details and testing details.

\subsection{Training Details}

Our experiments are conducted using PyTorch and Kornia, with model training performed on two A5000 GPUs. We train our model in three stages, with each stage consisting of 20, 14, and 20 epochs, respectively. During the second stage, we set the weight parameter of the loss functions, denoted as $\lambda_{1,2,3}$, to 1. In the third stage, we set $\lambda_{1}$ to 10, while $\lambda_{2,3,4}$ are set to 1. In the first stage, we set the batch size to 32, while in the second stage, we set it to 10. To accommodate the limited GPU memory, we utilize gradient accumulation in stages two and three, performing backpropagation after every two batches. We employ AdamW as the optimizer, with a learning rate of $10^{-3}$ for the first and second stages, and $10^{-4}$ for the third stage, while the weight decay is fixed at $10^{-2}$. We evaluate the performance of our model using different message extractors, including ResNet-50 \cite{he2016deep}, ConvNeXt-T \cite{liu2022convnet}, ViT-B \cite{dosovitskiy2020image}, and Swin-T \cite{liu2021swin}.

To account for different types of distortion combinations in each stage, we construct distinct distortion pipelines. Within each pipeline, each distortion is applied with a probability of 50\%.

In the first stage of our pipeline, we apply distortion to the pattern. Firstly, a random rectangular area is removed from the image and replaced with a zero value. The proportion of the erased area in the image is randomly selected from the range of [0.02, 0.33], while the aspect ratio of the erased area is randomly selected from the range of [0.3, 3.3]. Following this, we apply a perspective transformation with a scale range of [0, 0.7]. Finally, an affine transformation is applied with a rotation angle range of [-15$^\circ$, 15$^\circ$], scaling range of [1, 2], and translation range of [-0.3, 0.3] in both horizontal and vertical directions.

In the second stage of our pipeline, we apply spatial distortions, including perspective and affine transformations, to the encoded image. The scaling range of the affine transformation is [0.15, 1], while all other distortion ranges are the same as in the first stage. Pixel-wise distortions are then applied to the composite image. For color jitter, the factors of brightness, contrast, and saturation are randomly selected from the range of [0, 0.3], while the factor of hue is randomly selected from the range of [0, 0.1]. For Gaussian blur, the kernel size is 3$\times$3, and $\sigma$ is randomly selected from the range of [0.1, 1]. For motion blur, the kernel size is 3 and the direction is arbitrary. Gaussian blur is applied with a mean of 0 and a variance of 0.05. To address the non-differentiability of the DCT quantization process of image blocks in JPEG compression, we utilize a piece-wise function to simulate the quantization process, as described in the paper \cite{shin2017jpeg}, \ie,
\begin{equation}
\begin{aligned}
    q(x) = \begin{cases}
    x^{3}, &  \mid x \mid < 0.5; \\
    x,  &  \mid x \mid \geq 0.5. \\
    \end{cases}
\end{aligned}
\end{equation}
We randomly select the quality factor from the range of [50, 100].

\subsection{Testing Details}

\subsubsection{Setting of Experiment in the Wild}

In our study, we present encoded images at their default size, which is 7cm$\times$7cm on the monitor. However, on a laptop, the image dimension reduces to 5cm$\times$5cm. Following printing, the image dimension increase to 20cm$\times$20cm.

\subsubsection{Setting of Digital Distortion}

In Section 4.3 of the main manuscript, we conduct tests to evaluate the robustness of our model under various digital distortions. In this section, we provide a detailed explanation of the meaning of each abscissa used in our experiments. For Brightness, Contrast, Saturation, and Hue, the abscissa represents different factors. For Gaussian noise, we set the mean to 0, and the abscissa represents different standard deviations ($\sigma$). For JPEG compression, the abscissa represents different quality factors. For Gaussian blur, we set the standard deviation ($\sigma$) to 100, and the abscissa represents different kernel sizes. For perspective transformation, the abscissa represents different scales used to manipulate the perspective of the images. Finally, for translation, the abscissa represents the maximum absolute fraction used for horizontal and vertical translations.

In the main body's section 4.6, we incorporate combined distortions to the images by simultaneously applying various types of distortions. These distortions encompass brightness, contrast, saturation, hue, Gaussian blur, motion blur, Gaussian noise, JPEG compression, perspective transformation, and translation. The intensity of each distortion applied to the images during testing remain consistent with the strength employed during training.

\clearpage

\section{More Visual Examples}
In this section, we provide more visual examples, in terms of visual quality and robustness.

\subsection{Visual Quality}

\begin{figure}[h]
  \centering
  \includegraphics[width=1\linewidth]{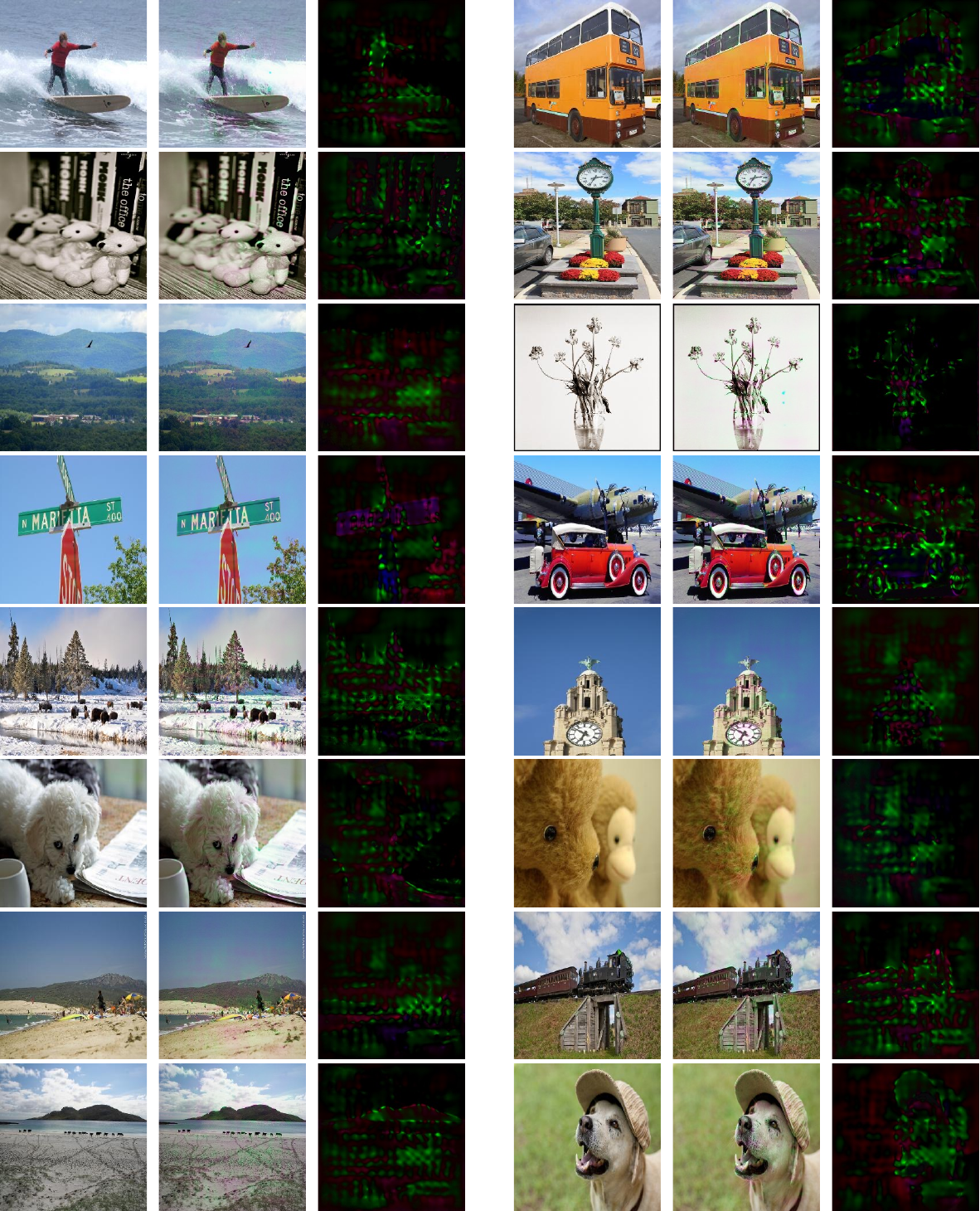}
  \caption{Visual examples of original images, encoded images, and their corresponding 5 $\times$ residuals.}
  \label{fig:encoded_image_samples}
\end{figure}

\Fref{fig:encoded_image_samples} presents a set of illustrative examples that demonstrate the complementary nature of visual quality, showcasing both encoded images and residuals.

\subsection{Robustness Against Digital Distortions}
\begin{figure}[h]
  \centering
  \includegraphics[width=0.7\linewidth]{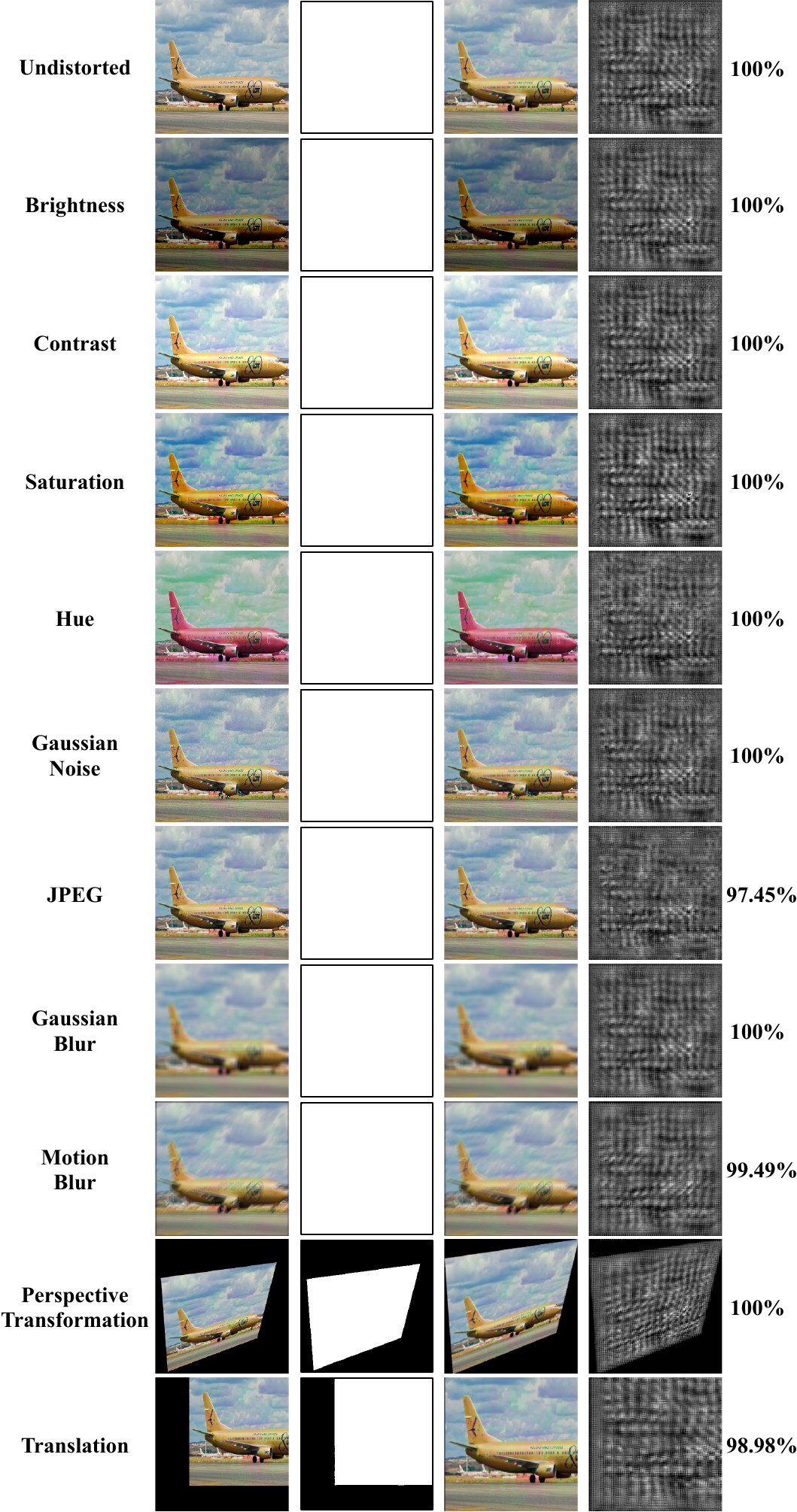}
  \caption{Robustness against digital distortions. Each column shows distorted images, localization masks, cropped images, decoded patterns, and the corresponding BER.}
  \label{fig:digital_distortions}
\end{figure}

\clearpage
\subsection{Robustness Against Physical Robustness }

\begin{figure}[h]
  \centering
  \includegraphics[width=0.8\linewidth]{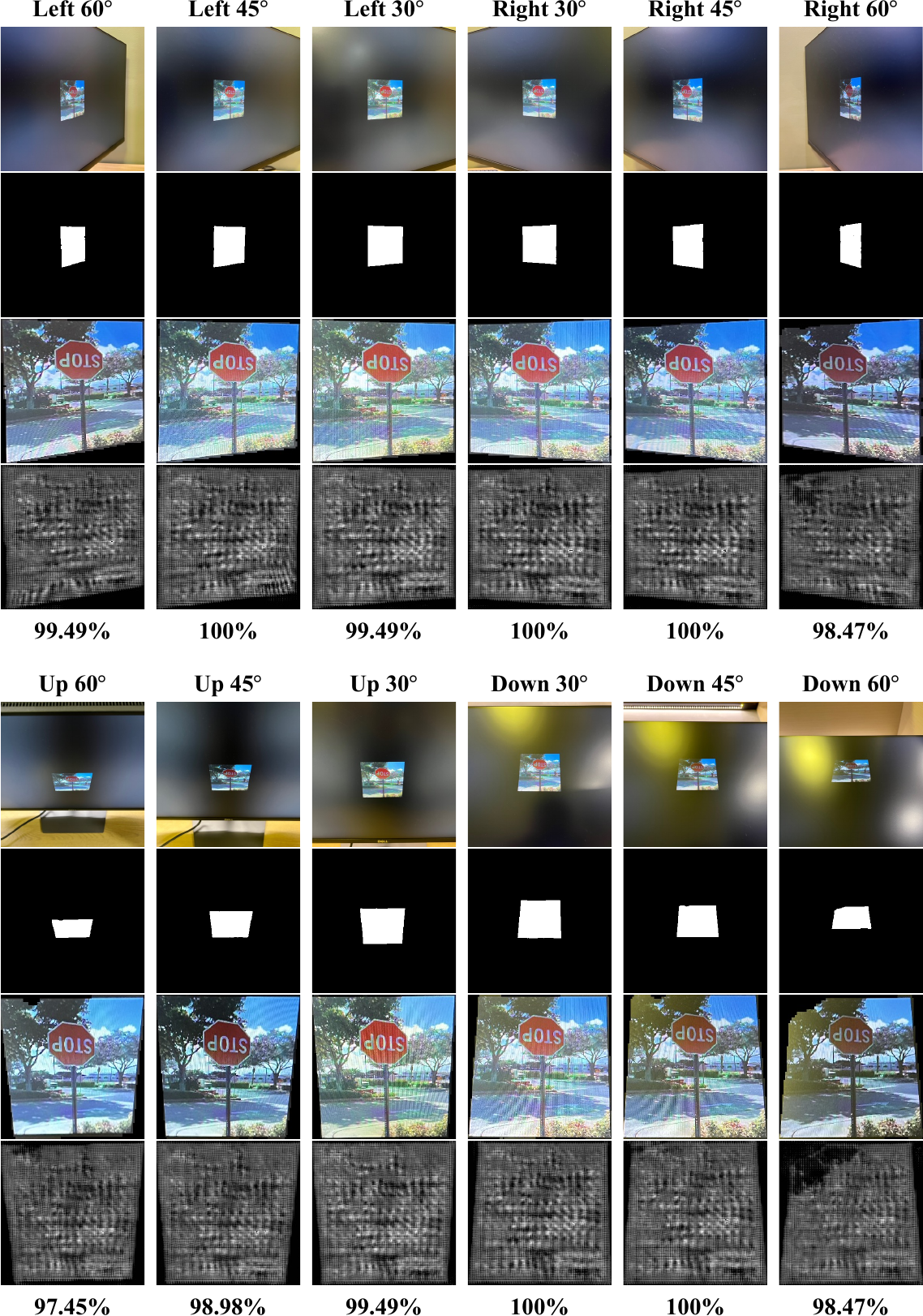}
  \caption{Visual illustrations of screen-shooting under different angles. Each row displays physical photos, localization masks, cropped images, decoded patterns, and the corresponding BER.}
  \label{fig:screen_shooting_angles}
\end{figure}

\begin{figure}[t]
  \centering
  \includegraphics[width=0.8\linewidth]{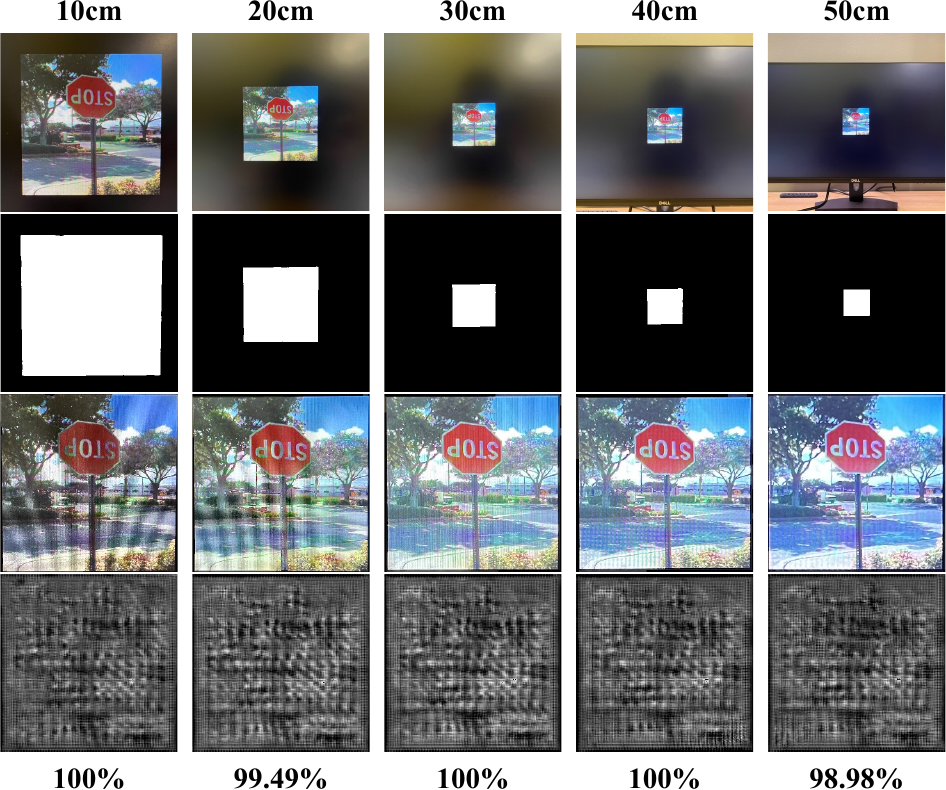}
  \caption{Visual illustrations of screen-shooting under different distances. 
  Each row displays physical photos, localization masks, cropped images, decoded patterns, and the corresponding BER.}
  \label{fig:screen_shooting_distances}
\end{figure}

\begin{figure}[t]
  \centering
  \includegraphics[width=0.8\linewidth]{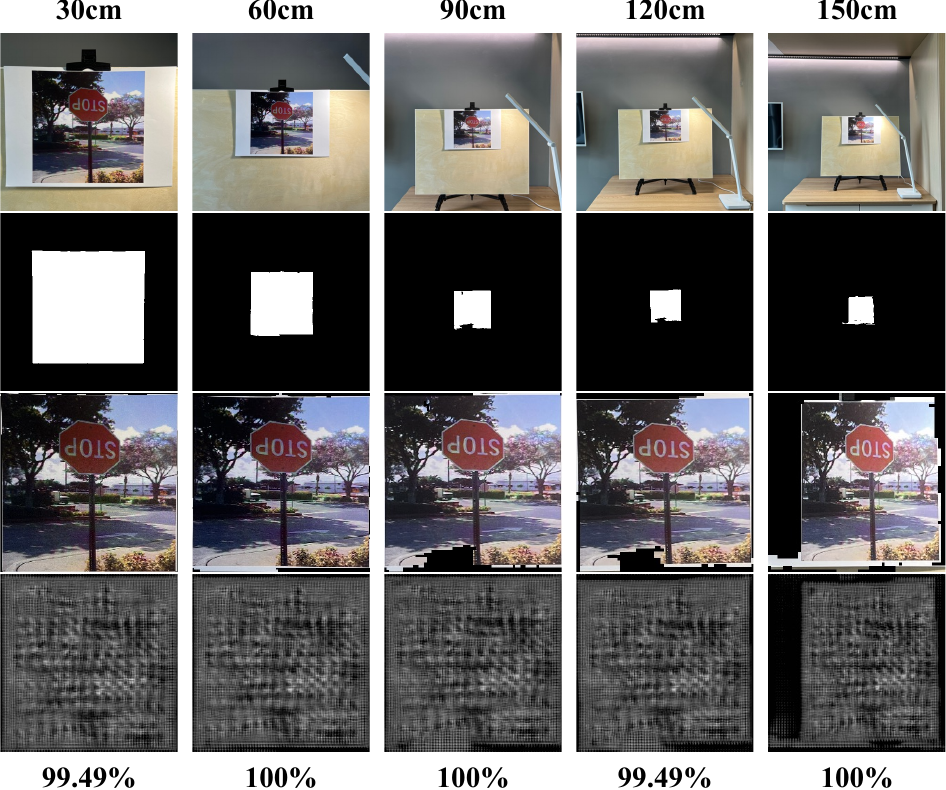}
  \caption{Visual illustrations of print-shooting under different distances.
    Each row displays physical photos, localization masks, cropped images, decoded patterns, and the corresponding BER.}
  \label{fig:print_shooting_distances}
\end{figure}

\begin{figure}[t]
  \centering
  \includegraphics[width=0.8\linewidth]{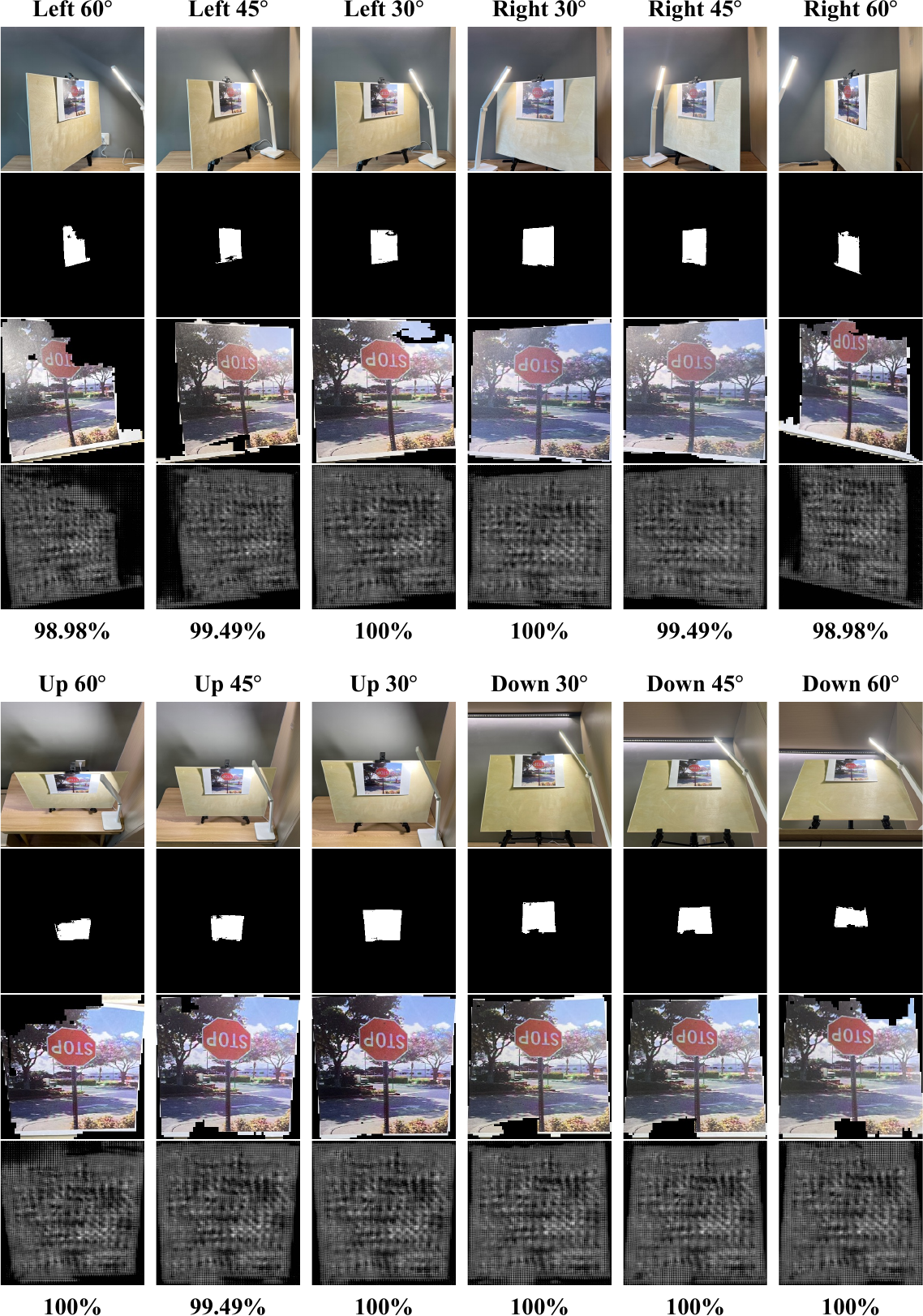}
  \caption{Visual illustrations of print-shooting under different angles.   Each row displays physical photos, localization masks, cropped images, decoded patterns, and the corresponding BER.}
  \label{fig:print_shooting_angles}
\end{figure}

We present visual examples of screen-shooting under different angles and distances in \Fref{fig:screen_shooting_angles} and \Fref{fig:screen_shooting_distances}, respectively. In addition, visual samples of print-shooting under different distances and angles are presented in \Fref{fig:print_shooting_distances} and \Fref{fig:print_shooting_angles}.

\clearpage

\section{Limitations}
\begin{figure}[h]
  \centering
  \includegraphics[width=0.5\linewidth]{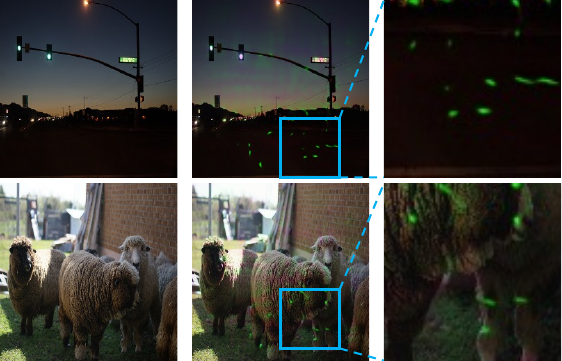}
  \caption{Examples of encoding for dark images. The 1st column is original images, the 2nd column is the encoded images, and the 3rd column shows zoomed-in images of the bright spots in the images.}
  \label{fig:dark_image}
\end{figure}

While our method achieves exceptional visual quality results, we empirically find that embedding messages into dark images may result in bright spots that compromise the aesthetic appeal, as depicted in the \Fref{fig:dark_image}.
We explain that dark images have low redundancy for watermarking.

\begin{figure}[h]
  \centering
  \includegraphics[width=0.9\linewidth]{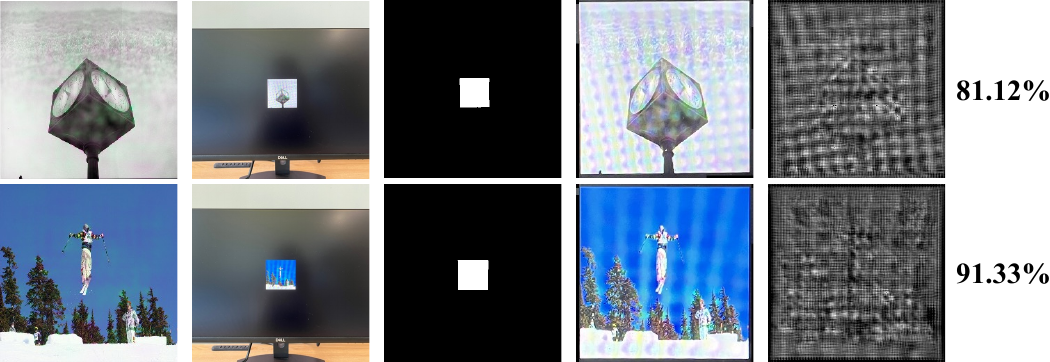}
  \caption{Examples of decoding under severe moire distortions. Each column shows original encoded images, physical photographs, localization masks, cropped images, decoded patterns, and the corresponding BER.}
  \label{fig:moire_pattern}
\end{figure}

Our method exhibits remarkable resilience against screen-shooting attacks. However, when confronted with pronounced moire patterns, the decoding performance may significantly deteriorate, as illustrated in the \Fref{fig:moire_pattern}.

\end{document}